\documentclass[aps,showpacs,twocolumn,floatfix,pre]{revtex4}

\usepackage{amsfonts}
\usepackage{amsmath}
\usepackage{amssymb}

\usepackage{graphicx}

\begin{document}
\large
 \title{
 Optimization of soliton ratchets in inhomogeneous sine-Gordon systems }
         
\author{F. G. Mertens$^1$, L. Morales-Molina$^2$, A. R. Bishop$^3$, A. S\'anchez$^{4}$ and P. M\"uller$^1$ }
\affiliation 
{$^1$Physikalisches Institut, Universit\"at Bayreuth, D-95440 Bayreuth,
Germany}

 \affiliation 
{$^2$Max-Planck-Institut f\"ur Physik komplexer Systeme, N\"othnitzer Str. 38, D-01187 Dresden, Germany}

\affiliation 
{$^3$Theoretical Division and Center for Nonlinear Studies, 
Los Alamos National Laboratory,  Los Alamos, NM 87545, USA}
\affiliation 
{$^4$Grupo Interdisciplinar de Sistemas Complejos (GISC) 
Departamento de Matem\'aticas,
Universidad Carlos III de Madrid, Avenida de la Universidad 30, 28911
Legan\'es, Madrid, Spain }
\affiliation 
{Instituto de Biocomputaci\'on
y F\'isica de Sistemas Complejos (BIFI), Universidad de Zaragoza,
50009 Zaragoza, Spain}

\date{today}

\pacs{05.45.Yv, 
85.25.Cp, 
73.40.Ei 
}

\begin{abstract}
Unidirectional motion of solitons can take place, although the applied
force has zero average in time, when the spatial symmetry  
is broken by introducing a potential
$V(x)$, which consists of periodically repeated cells with each cell containing an
asymmetric array of strongly localized inhomogeneities at positions
$x_{i}$. A collective coordinate approach shows that the positions, heights and
widths of the inhomogeneities (in that order) are the crucial parameters so as
to
obtain an optimal effective potential $U_{opt}$ that yields a maximal average soliton
velocity. $U_{opt}$ essentially exhibits two features:
double peaks consisting of a positive and a negative peak, and long flat regions
between the double peaks. Such a potential can be obtained by choosing
inhomogeneities with opposite signs (e.g., microresistors and microshorts in the
case of long Josephson junctions) that are positioned close to each other, while the
distance  between each peak pair is rather large. These results of the
collective variables theory are confirmed by full simulations for the inhomogeneous
sine-Gordon system.

\end{abstract}

\maketitle

\section{Introduction}

Ratchet or rectification phenomena appear in many different fields
ranging
from nanodevices to molecular biology 
\cite{Hanggi, Astumian, Ajdari, Reimann, Linke}. 
In the simplest model
a point-like particle is considered which is driven by deterministic or
non-white stochastic forces. Under certain conditions related to the
breaking of symmetries, unidirectional motion of the particle can take
place
although the applied force has zero average in time.

These particle ratchets have been generalized to spatially
extended nonlinear systems, in which solitons play a similar role to the
above point particles \cite{Zapata, Marchesoni, falo, trias, salerno2, const, gorbach}. 
In particular, solitons in nonlinear Klein-Gordon systems
have been shown to move on the average in one direction, although the
driving force has zero time average, if either a temporal or a spatial
symmetry is broken.

In the first case, a biharmonic driving force has been used which breaks
a
time shift symmetry \cite{flach, Salerno-mixing}. Here the mechanism of the ratchet effect has
been
clarified in detail by a collective variable theory \cite{luisprl, luis05}, which uses
the
soliton position $ X(t)$ and width $l(t)$. Due to the coupling between the
translational and internal degrees of freedom, energy is pumped
inhomogeneously into the system, generating  a directional
motion.
The breaking of the time shift symmetry gives rise to a resonance
mechanism
that takes place whenever the width $l(t)$ oscillates with at least one
frequency of the external ac force. This ratchet effect has been
confirmed
by experiments with annular Josephson junctions \cite{JJ2} which can be modeled
by sine-Gordon systems; here flux quanta (fluxons) play the role of the
solitons. As the external ac force, biharmonic microwaves have been used. 

The biharmonic force has
recently been replaced by more general periodic forces which can be
expressed
by Jacobian elliptic functions. There, it turns out that the average soliton velocity 
exhibits an extremum for a  certain value of the modulus of the elliptic functions  \cite{Chacon}. 
This means that there is an optimal form for the periodic force.

Another way to obtain a soliton ratchet is to break the spatial
symmetry.
This has been demonstrated very recently by the introduction of
point-like inhomogeneities into nonlinear Klein-Gordon systems \cite{luis04, luispre}.
These
inhomogeneities can be modeled by $\delta $-functions, if their spatial
extent
is much smaller than the characteristic length for the system (the
Josephson
penetration length in the case of the Josephson junctions).
In order to achieve a ratchet effect the $\delta $-functions must form
asymmetric
periodic arrays. This asymmetry translates into an asymmetry of an
effective
potential $U(X,l)$ which appears in a collective variable theory with the
variables $X(t)$ and $l(t)$.

As an example, three $\delta$-functions of equal strength have been
positioned in each cell of an array with period $L$. The above theory
yielded
a nearly perfect agreement with the simulations for the inhomogeneous
sine-Gordon system \cite{luispre}. By choosing different arrangements for the
positions of the $\delta $-functions within the cells of the array it was
possible to increase the average soliton velocity $\langle v \rangle$ which means that
the
ratchet system provides better transport.

The aim of this paper is to search for the optimal shape and arrangement
of localized inhomogeneities, in the sense that $\langle v \rangle$ becomes as large as
possible. This means that instead of $\delta$-functions other functions
which
represent localized inhomogeneities (e.g., Gaussians, Lorentzians,
box-type functions etc.) have to be tested, and that the arrangement of
these inhomogeneities within the cells, but also the length $L$ of the
cells
in the periodic array, all have to be optimized.

In section II nonlinear Klein-Gordon systems with localized inhomogeneities
are introduced. In section III a collective coordinate (CC) theory is 
developed which results in a set of ODEs containing an effective potential $U$
which depends on the inhomogeneities. It turns out that their most significant
features for maximizing $\langle v \rangle$ are their positions, heights and widths, while their detailed shape
is unimportant. For this reason we work with the simplest shape, namely that
of boxes, where $U$ can be calculated analytically. In section IV the optimal
effective potential $U_{opt}$ is obtained by using an ansatz in terms of
Jacobi elliptic functions. $U_{opt}$ essentially  exhibits two features:
double peaks consisting of a positive and a negative peak with a steep 
slope in between, and long flat regions between the double peaks. In section V
we show that the essential features of $U_{opt}$ can be produced by choosing 
appropriately the positions, heights and widths of box inhomogeneities. In 
this way very high soliton velocities can be obtained. These results of the
CC- theory are confirmed by full simulations for the sine-Gordon system with the
above inhomogeneities  (section VI). 

In the last section we stress 
that an important step in the optimization consists
 in using inhomogeneities with {\em opposite} signs. 
Both types of inhomogeneities were already used in long Josephson junctions, namely microresistors (critical current $J_c$ 
decreased) and microshorts ($J_c$ increased) \cite{usti}. However, so far only equidistant arrays of either type were used in 
experiments. We propose to use a specific combination of  both types, resulting in an asymmetric periodic array, which 
yields very high average soliton velocities both in a collective coordinate
theory and in full simulations for the inhomogeneous 
sine-Gordon system.\newline

\section{Inhomogeneous nonlinear Klein-Gordon systems}

A ratchet effect has recently been obtained \cite{luis04, luispre} by breaking the spatial symmetry of nonlinear Klein-Gordon 
systems with inhomogeneities introduced via a potential $V (x)$
\begin{equation}
\label{aa}
\phi _{tt} + \beta \phi _t - \phi _{xx} + \frac{\partial \tilde{U}}{\partial \phi } \, [1 + V(x)] = f(t).
\end{equation}

Here $\phi  (x, t)$ is a scalar field, $\phi _x$ and $\phi _t$ are partial derivatives with respect to space and time, $\beta $ is a 
damping coefficient,
$\tilde{U}(\phi ) = 1 - \cos \phi $ is the sine-Gordon (sG) potential , and $\tilde{U} = \frac{1}{4} (\phi ^2 -1)^2$ 
the $\phi ^4$-potential. As both cases have produced
very similar results, \cite{luispre} we will concentrate on the sG-model in the following, but we will always indicate which results
also hold for the $\phi ^4$-model. $f(t) = A \sin (\omega t + \delta _0)$ is an external ac force with amplitude $A$, frequency $\omega $ and initial
phase $\delta _0$.

$V(x)$ consists of periodically repeated cells of length $L$; each cell $n$ contains an asymmetric array of strongly localized inhomogeneities 
$g_i(x)$ which are placed at positions $x_i$ within the cell; i.e. 

\begin{equation}
\label{ab}
V(x) = \sum_n \sum_i \, g_i (x - x_i - nL).
\end{equation}

So far only the case of point-like inhomogeneities was considered which can be modeled by $\delta $-functions. Taking three
 $\delta $-functions with equal strengths and choosing their positions $x_i$ in certain asymmetric ways, a sine-Gordon soliton 
(also named kink in the following) moves either to the right or to the left on
 the average \cite{luis04}. Depending on the choice of the driving frequency
 $\omega$, qualitatively different results were obtained for the modulus
 $\bar{v}$ of the average soliton speed $\langle v \rangle$ as a function of
 the driving amplitude $A$: a) For the low frequency $\omega=0.015$ there is
 an up and down staircase with the maximum average velocity $v_{max}=0.03820$,
 b) for the intermediate frequency $\omega=0.05$ there are five 
 "windows" (regions of $A$ with constant $\bar{v}>0$ ) with 
$v_{max}$=0.03183, c) for the relatively high frequency $\omega=0.1$
 there is only one window with height $v_{max}=0.063666$. For even higher
 frequencies, $\bar{v}$ vanishes for all amplitudes $A$. The results for
 $\phi^4$ nonlinearity are quite similar. 

$v_{max}$ is always small compared to the critical velocity $c$ which is unity
 for the dimensionless system (\ref{aa}). For long Josephson junctions $c$
 is the Swihart velocity \cite{waves}. 

The aim of this paper is to optimize both the shape and the array of the inhomogeneities in the sense that $\bar{v}$ becomes as large as possible. Naturally, there are still other criteria for the optimization. E.g., one can try to maximize
 the area under the curve $\bar{v}$ 
vs. amplitude $A$ ; this will be discussed later. Other optimization
 strategies include studying the energy efficiency. We do not
 consider this approach here, but we refer the reader to Ref.\cite{parrondo}.

In order to achieve any kind of optimization, it is necessary to have a theory
which allows calculation of $\langle v \rangle$ for a given array 
of inhomogeneities. A CC theory with two variables, namely position $X(t)$ and width $l(t)$ of the soliton, 
has turned out to be very successful \cite{luis04}, in the sense that in the
case of $\delta $-functions the results for $\langle v \rangle$ as a 
function of the driving amplitude $A$ agree very well with the simulations for
the original system (full numerical solution of 
Eq. (\ref{aa})). Deviations between theory and simulation are found only in the case of strong driving $(A = {\cal O} (1))$. For even stronger driving the theory is 
no longer applicable, because in the simulations kink-antikink pairs appear spontaneously and scatter with the soliton under consideration.
The 2-CC theory is characterized by two ODEs for $X(t)$ and $l(t)$, which contain forces that are defined as partial derivatives of an
effective potential $U(X, l)$, which is calculated and discussed in the next
section. 

We take for our computations a rather large damping, choosing $\beta=1$, for the following reasons:\newline
a) Our CC-theory does not take into account the phonons which are radiated 
by the soliton due to its acceleration by the driving force. These phonons
are quickly damped out when the damping is large.\newline
b) We want to avoid several phenomena which can occur for very small damping:
 e.g., chaotic behavior and current reversals.
   
Finally, we remark that the spatial symmetry can be broken also by the
introduction of an additive inhomogeneity $\gamma(x)$ in Eq.\ref{aa}, instead
of our multiplicative inhomogeneity $\frac{\textstyle \partial
  \tilde{U}}{\textstyle \partial \phi } V(x)$. In the case of long Josephson junctions
an additive inhomogeneity has been experimentally realized by a current
injection with a profile $\gamma(x)$ \cite{goldobin}. This produces an
effective potential which is similar to an asymmetric potential.

\section{Collective coordinate theory}

For the unperturbed sine-Gordon equation, i.e. Eq. (\ref{aa}) without damping, inhomogeneity and driving, the 1-soliton solution reads

\begin{equation}
\label{ac}
\phi _0 (x, t) = 4 \arctan \bigl [\exp (\gamma \frac{x - vt}{l_0}) \bigr ],
\end{equation}
where $v$ is the soliton velocity, $l_0 = 1$ is the rest width, and $l_0/\gamma = l_0  \sqrt{1 - v^2}$ is the Lorentz contracted width.

For perturbed nonlinear Klein-Gordon systems the so-called Rice ansatz \cite{rice} turned out to be very successful. E.g., for the ac 
driven $\phi ^4$-system an unexpected resonance was predicted (and confirmed by simulations), which is situated in the gap 
below the phonon 
spectrum at half the frequency of the internal mode of the $\phi ^4$-kink \cite{quintprl, quintpre}. The Rice ansatz for the sG-systems reads 

\begin{equation}
\label{ad}
\phi  (x, X, l) = 4 \arctan \{  \exp \bigl [ \frac{x - X(t)}{l (t)} \bigr ] \}
\end{equation}
with soliton position $X(t)$ and width $l(t)$. The 2-CC theory for the perturbed sG-equation (\ref{aa}) yielded a set of two coupled ODEs 
\cite{luispre}

\begin{equation}
\label{ae}
M_0 l_0 \frac{\ddot{X}}{l} + \beta M_0 l_0 \frac{\dot{X}} {l} \,- \, M_0 l_0 \frac{\dot{X} \dot{l}} {l^2} = F^{ac} + F^{inh}
\end{equation}

\begin{equation}
\label{af}
\alpha M_0 l_0 \frac{\ddot{l}}{l} + \beta \alpha  M_0 l_0 \frac{\dot{l}} {l} \, + \, M_0 l_0 \frac{\dot{X}^2} {l^2}= K^{int} + K^{inh},
\end{equation}
where $\alpha  = \pi ^2/12, M_0 = 8, l_0 = 1$ and 
\begin{equation}
\label{ag}
F^{ac} = \int_{- \infty}^{\infty} dx f (t) \frac{\partial \phi }{\partial X} = - q f (t)
\end{equation}
with the topological charge $q = 2 \pi $.

\begin{equation}
\label{ah}
K^{int} = - \frac{\partial E}{\partial l}
\end{equation}
is a force which arises from the soliton energy
\begin{equation}
\label{ai}
E(\dot{X},  l,  \dot{l}) = \frac{1}{2} \, \frac{l_0}{l} M_0 \dot{X}^2 + \frac{1}{2}  \frac{l_0}{l} \alpha   M_0 \dot{l}^2 +  \frac{1}{2} M_0 (\frac{l_0}{l} + \frac{l}{l_0}),
\end{equation}
including internal energy due to width oscillations.

There are two forces which appear due to the potential $V(x)$ in
Eq. (\ref{aa}), viz.

\begin{eqnarray}
\label{aj}
F^{inh}& =& - \int_{- \infty}^{\infty} dx  \frac{\partial \tilde{U} }{\partial
  \phi } \frac{\partial \phi }{\partial X} V(x) 
\nonumber \\
&=& 
- \int_{- \infty}^{\infty} dx \frac{\partial \tilde{U} }{\partial X } V(x): = -  \frac{\partial U }{\partial X}
\end{eqnarray}

\begin{eqnarray}
\label{ak}
K^{inh}& =& - \int_{- \infty}^{\infty} dx  \frac{\partial \tilde{U} }{\partial
  \phi } \frac{\partial \phi }{\partial l} V(x) 
\nonumber \\
&=& - \int_{- \infty}^{\infty} dx \frac{\partial \tilde{U} }{\partial l } V(x): = -  \frac{\partial U }{\partial l} \,.
\end{eqnarray}
In this way the effective potential 

\begin{equation}
\label{al}
U(X, l) = \int_{- \infty}^{\infty} dx \tilde{U} (\phi ) V(x)
\end{equation}
is introduced that characterizes the influence of the inhomogeneities on the soliton dynamics.
We remark that all the above results also hold for the $\phi ^4$-model:  In
  this case  $\phi  = \tanh [(x - X)/l]$ and the parameters are
$q = 2, M_0 = 2 \sqrt{2}/3, l_0 = \sqrt{2}$, and $\alpha  = (\pi ^2 - 6)/12$.

Before we specify $V(x)$ we evaluate $\tilde{U}(\phi)$. First we derive the
relation $\tilde{U}(\phi _0) = (\frac{\partial \phi _0}{\partial
  z})^2/(2\gamma ^2) = 2 \, \text{sech}^2 (\gamma  z)$, with $z = x - vt, $ by
inserting $\phi _0$ into the unperturbed sG-equation. 
However,  $\tilde{U}(\phi)$ can not be obtained in this way because 
$\phi(x,X,l)$ is not an exact solution, but rather an ansatz. However, we can generalize 
the above relation  for $\tilde{U}(\phi _0)$ to $\tilde{U} (\phi ) =
 \frac{1}{2} l^2 \phi _z^2 = 2 \,\text{sech}^2 (z/l)$, with $z = x - X$, and
 this can be verified by e.g.
 MATHEMATICA \cite{wolfram}.

Thus we finally obtain

\begin{equation}
\label{am}
U (X, l) =  \int_{- \infty}^{\infty} dx \, 2 \, \mbox{sech}^2 \frac{x-X}{l} \cdot V(x) .
\end{equation}

Now we can insert the superposition (\ref{ab}) of localized inhomogeneities $g_i$ and obtain 

\begin{equation}
\label{an}
U (X, l) = \sum_n \sum_i U_i^{(n)}
\end{equation}

\begin{equation}
\label{ao}
U_i^{(n)} =  \int_{- \infty}^{\infty} dx \, 2 \, \mbox{sech}^2 \frac{x-X}{l} \cdot g_i (x - x_i - n L).
\end{equation}
We can check that the effective potential is indeed periodic, $U(X + L, l) = U (X, l)$, because the sum goes over an infinite number of unit 
cells $n$ of length $L$. 

Our next step is to evaluate $U_i^{(n)}$ for various strongly localized functions $g_i$ and to discuss which are the most important features 
of the inhomogeneities. We have tested Gaussians, Lorentzians and box functions and it turns out that there are only two important 
features of $g_i$, namely the height and width. For the above three cases, the $U_i^{(n)}$ are always bell-shaped, and the desired asymmetry 
of $U$ depends most strongly on the relative positions of the $U_i^{(n)}$ within the cells. In other words, the most important features 
of the inhomogeneities $g_i$ are their positions $x_i$, then come their heights and widths, while their detailed shape is unimportant. 

For this reason, we can choose the simplest case, namely the box functions, which has the additional advantage that $U(X, l)$ can 
be calculated analytically (in contrast to the other two cases). The box function $g_i$ at $x_i$ in cell $n = 0$ is defined as 

\begin{equation}
\label{ap}
g_i = \begin{cases} h_i & \text{for} \quad x_i - b_i \leq x \leq x_i + b_i \\ 
0 & \text{otherwise} \end{cases}\, .
\end{equation}

This yields

\begin{eqnarray}
\label{aq}
U_i^{(o)} = 2 \,h_i\, l\, \sinh \frac{2 b_i}{l}\,\ \mbox{sech} \frac{X - x_i + b_i}{l} \nonumber \\
\times\,\ \mbox{sech} \frac{X - x_i - b_i}{l} 
\end{eqnarray}
and we can check that in the limit $h_i \rightarrow \infty, b_i \rightarrow 0$ with finite $\epsilon _i: = 2 b_i h_i$ the result for a $\delta $-function 
inhomogeneity \cite{luis04} can be regained: $U_i^{(0)} = 2 \, \epsilon _i \, \mbox{sech} ^2 [(X - x_i)/l]$. The final result for the effective 
potential (\ref{an}) is

\begin{equation}
\label{ar}
U(X, l) = \sum_n \sum_i 2 \, h_i\, l\, \sinh \frac{2 b_i}{l} \, \mbox{sech} \frac{Z_+}{l} \, \mbox{sech} \frac{Z_-}{l}
\end{equation}
with $Z_{\pm} = X - x_i - n L \pm b_i$.

In this way the optimization problem is solved, but only in principle: We choose a set of box inhomogeneities at 
the positions $x_i$ within the cells, numerically solve the ODEs (\ref{ae}), (\ref{af}) for $X(t), l(t)$, and compute the average soliton 
velocity, defined by 

\begin{equation}
\label{as}
\langle v \rangle =\langle \dot{X} \rangle  =  \lim_{t\to\infty} \frac{\langle
  X(t) - X(0) \rangle }{t}\, ,
\end{equation}
choosing a sufficiently long integration time. Then the procedure has to be repeated for many values of the 
driving amplitude $A$ (keeping the driving frequency $\omega $ fixed), in order to find the maximal soliton speed

\begin{equation}
\label{at}
 v_{max} = Max \{|\langle v \rangle |\}
\end{equation}
for that set of parameters. Finally, we would like to obtain $v_{max}$ for a low, a medium and a relatively high frequency, 
because it is known from the case of the $\delta $-function inhomogeneities
that the curves $\langle v \rangle$ vs. $A$ differ 
qualitatively for the above choice of frequencies (see section II).

In practice the whole procedure cannot work because the dimension of the parameter space is far too high. E.g., if one takes three 
inhomogeneities per cell, as in the case of the $\delta $-functions, one has 10 parameters, namely three positions, heights and widths 
of the boxes, plus the cell length $L$. As one has to make sweeps in $A$ in order to get $v_{max} $ for every parameter set, it is 
practically impossible to explore the 10-dimensional parameter
space. Obviously, we have to find a more efficient way to obtain 
an optimal set of inhomogeneities.

\section{Optimal effective potential}

Our approach is not to work directly with the potential $V(x)$ in the inhomogeneous sG-equation (\ref{aa}), but to design first an optimal 
\texttt{effective} potential $U_{opt}$ for a given driving force, i.e. for fixed $A$ and $\omega $. When $U_{opt}$ has been obtained, 
it can be represented 
(approximately) by the superposition (\ref{ar}) of the pulse-shaped contributions $U_i^{(n)}$.  I.e., the many parameters 
in Eq. (\ref{ar}) could be determined for instance by a least-square fit of $U$ to $U_{opt}$, which is much more efficient than a search in the
 high-dimensional parameter space.

For simplicity we will not try to find $U_{opt} (X, l)$ within the 2-CC approach. It will turn out to be sufficient to find an optimal potential within the 
1-CC theory which starts with the ansatz 

\begin{equation}
\label{au}
\phi _0 (x, X) = 4 \arctan \{\exp [\gamma (x - X(t)]\} 
\end{equation}
with
\begin{equation}
\label{av}
\gamma  = 1/ \sqrt{1 - \dot{X}^2} .
\end{equation}
This yields one ODE \cite{luispre}

\begin{equation}
\label{aw}
\gamma ^3 M_o \ddot{X} + \gamma \beta \dot{X} = F^{ac} + F^{inh} \, ,
\end{equation}
with $F^{inh} = - \partial U/\partial X$ and 
\begin{eqnarray}
\label{ax}
U &=& \int_{- \infty}^{\infty} dx \tilde{U} (\phi ) V(x) \nonumber \\ 
&=& \int_{- \infty}^{\infty} dx \,2\, \mbox{sech}^2 [\gamma (x - X)] \, V(x).
\end{eqnarray}
Here, $\gamma ^3 M_0$ is the so-called longitudinal relativistic mass. So far only the non-relativistic limit 
$(\gamma  = 1)$ of Eq. (\ref{aw}) was considered in the context of the $\delta $-function inhomogeneities 
\cite{luis04, luispre}. Here we need the relativistic version, because our optimization will yield velocities which 
are no longer small compared to the critical velocity $c = 1$.

As noted above, we do not consider here the expression (\ref{ax}) for $U$ in terms of $V(x)$, but we want to find 
quite generally the optimal $U$ for a given driving force $F^{ac}$. $U_{opt}$ must be asymmetric and periodic. 
As the most general periodic functions are the Jacobi elliptic functions, we use the ansatz 

\begin{equation}
\label{ay}
U_{opt}(X) = - \epsilon \cdot sn (kX, m) \cdot cn (kX, m),
\end{equation}
where $m$ is the modulus of the elliptic functions, $ k = 2 K (m)/L$ is a generalized wave number and $K(m)$ 
is the complete elliptic integral of the first kind.

This type of ansatz was first introduced in  \cite{Chacon} as a generalized driving force to break the
 temporal symmetry in ratchet systems, namely
 $f(t) \equiv F_{ellip} (t) =
 \epsilon \cdot sn (\Omega t, m) \cdot cn ( \Omega t,  m)$ (where $\Omega  = 2 K (m)/T$ with  period $T$). 
Using only symmetry arguments, and taking into account the first two terms of the Fourier series of $F_{ellip} (t) $, it was 
shown in \cite{Chacon} that the optimal value for $m$ is $0.960057$, independent of the details of the ratchet 
models. However, in our case we will show that the situation is more
complicated when such an ansatz is taken for the potential $U(X)$ in the
 particle ratchet (\ref{aw}).
 
The negative sign in front of $\epsilon > 0$ in (\ref{ay}) is chosen in order 
to achieve a positive $\langle v \rangle$, i.e. average motion to the right.
The specific combination (\ref{ay}) of elliptic functions is chosen because there is no ratchet effect in two limiting cases: 
For $ m = 0$ the effective potential is sinusoidal, and for $m = 1$ it is zero (except for a set of points with Lebesgue 
measure zero). Thus we can expect that an optimal value $m_{opt}$ naturally exists in the interval $0 < m < 1$,  under the 
condition that there is any ratchet effect for the chosen set of parameters $\epsilon, L $ in $U(X)$ and $A, \omega $ 
in the driving force. 

\begin{figure}
\includegraphics[width=8.cm,height=8cm]{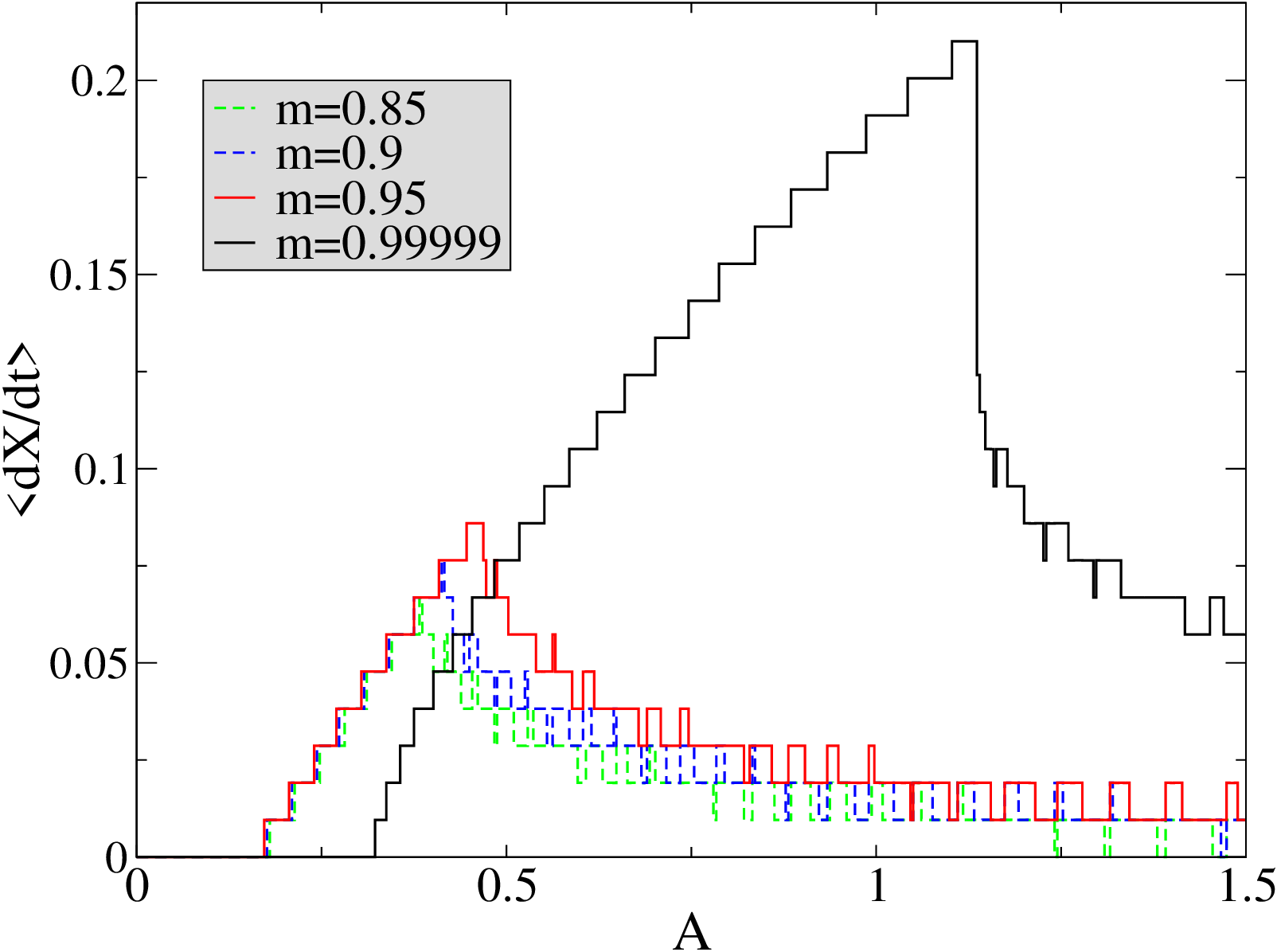}
\caption{Mean soliton velocity vs. amplitude  $A$ of the ac force for different moduli
  $m$ of
  the elliptic functions in the ansatz (\ref{ay}) for the effective potential. The parameters used
  are  $L=4$, $\beta=1$, $\omega=0.015$ and $\epsilon=-2$. }
\label{Fig1}
\end{figure}

\begin{figure}
\includegraphics[width=8.cm,height=8cm]{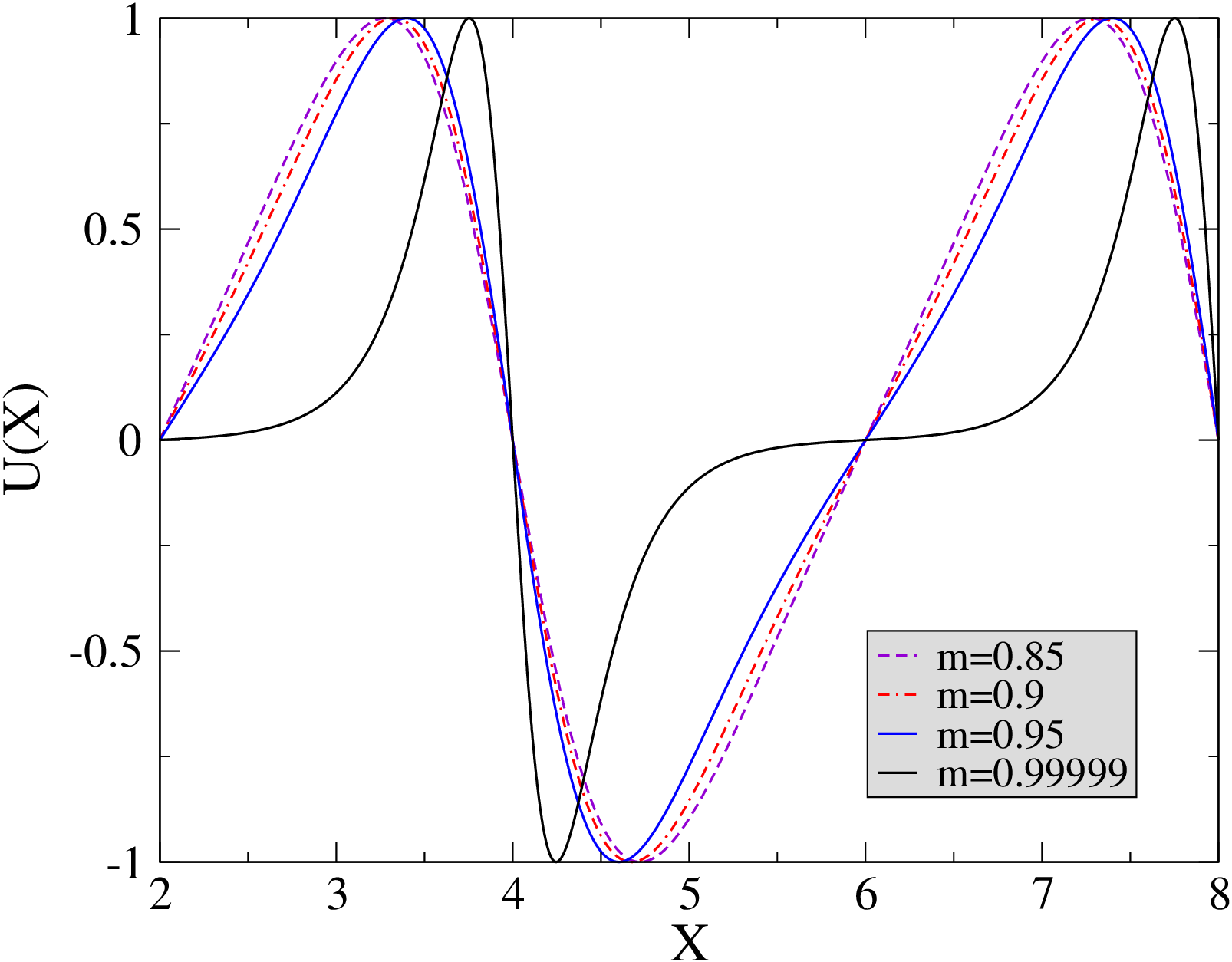}
\caption{Effective potential (\ref{ay}) for different values of the modulus
  $m$ of the elliptic functions; $L=4$, $\epsilon=-2$.}
\label{Fig2}
\end{figure}

 Our problem now is to find the optimal modulus $m$, i.e. the optimal shape of the potential $U(X)$ in the 
ODE (\ref{aw}), which formally is a relativistic particle ratchet model. In
Fig. \ref{Fig1} $\langle v \rangle $ vs. $A$ is plotted for various 
values of $m$ and we can see that the maximal speed $v_{max} $ gradually increases with increasing 
$m$. However, when $m$ is close to one, there is a dramatic increase of $v_{max} $. The reason for this can be discussed 
by considering the change of shape of $U(X)$ in Fig. \ref{Fig2}: When $m$ is increased (but is not yet close to 1) the negative 
slopes in $U(X)$ become steeper and the positive slopes become weaker. Both together increases the asymmetry of the potential; 
thus the symmetry breaking is augmented, which explains why $v_{max} $ grows. In fact, $U(X)$ looks like a (smoothed) sawtooth potential 
that has widely been used in the literature.

However, when $m$ becomes close to one, a new feature appears in $U(X)$, namely long flat regions. Their lengths increase as $m$ gets even 
closer to one, so eventually $U(X)$ essentially exhibits two features: \\
1.) double peaks, each consisting of a positive and a negative peak with a steep slope in between,\\
2.) long flat regions between the double peaks.

A potential with similar features was
proposed for a ratchet system driven by a multiplicative white noise
to enhance the coherent transport \cite{lindner} .

Before we proceed with our goal, namely to find the optimal inhomogeneities which produce the above 
features, we want to discuss the staircase structure of 
$\langle v \rangle$ vs. $A$ in Fig. 1. This structure is described by 
\begin{equation}
\label{az}
\langle v \rangle = \frac{i}{j}\, \frac{L}{T}  = \frac {i}{j} \,v_{step} \,;\, i, j: \,\mbox{integer} .
\end{equation}
The step height $v_{step} = 
L/T$ does not depend on the shape, but only on the period of $U(X)$, and on the period $T = 2 \pi /\omega$ of the ac force.

\begin{figure}
\includegraphics[width=8.cm,height=8cm]{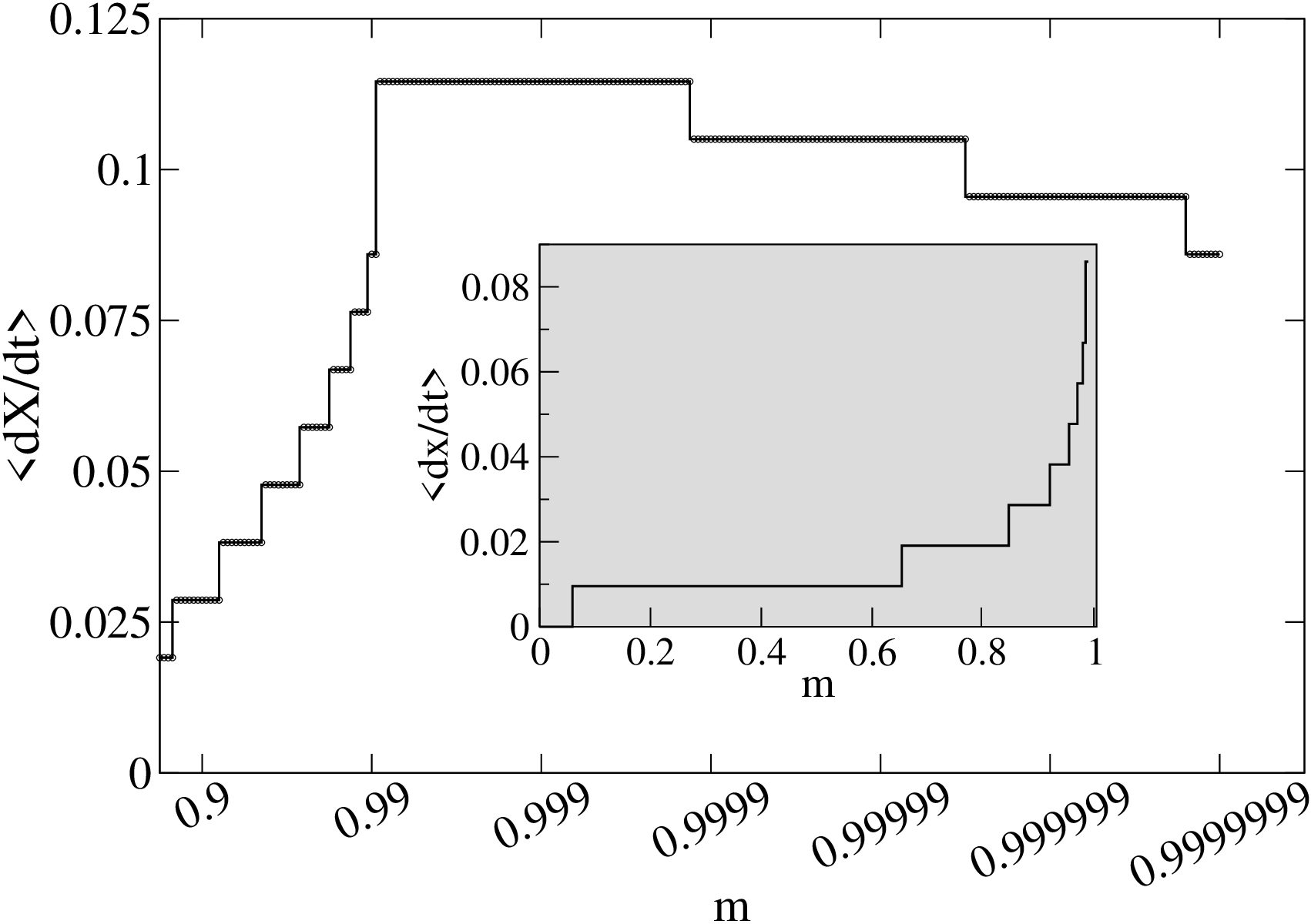}
\caption{Mean soliton  velocity vs. the modulus $m$ of the elliptic functions in
  the ansatz (\ref{ay}) for the effective potential; $L=4$,
  $A=0.6$, $\beta=1$, $\omega=0.015$, $\epsilon=-2$. }
\label{Fig3}
\end{figure}

Formula (\ref{az}) should hold quite generally according to most of the
literature ( \cite{Reimann} and references therein). However, Eq. (\ref{az}) was explicitly 
calculated only for a special case, namely an asymmetric sawtooth potential and a piecewise constant driving force  
 \cite{schreier}. Our results confirm Eq. (\ref{az}), and so do the results for the effective potential stemming from $\delta $-function 
inhomogeneities \cite{luis04}. Moreover, we observe that $\langle v \rangle $ changes in steps also as a function of other parameters. 
For the following it will be important
that $\langle v \rangle$  as a function of $m$ first increases, then decreases in steps (Fig. \ref{Fig3}). This means that 
  $v_{max}$ is reached for a very narrow range of $m$, which we denote by
$[m_{opt}]$. The larger $A$, the closer  
$[m_{opt}]$ is to $1$.  For a  larger value of the frequency,
  $\omega=0.05$, the same effect shows up, namely it is found that the ratchet
  transport is enhanced for $m$ very close to
  $1$ provided that $A$ is not too small.

\section{Optimal inhomogeneities}

\begin{figure}
\begin{center}
\includegraphics[width=8.cm,height=8cm]
{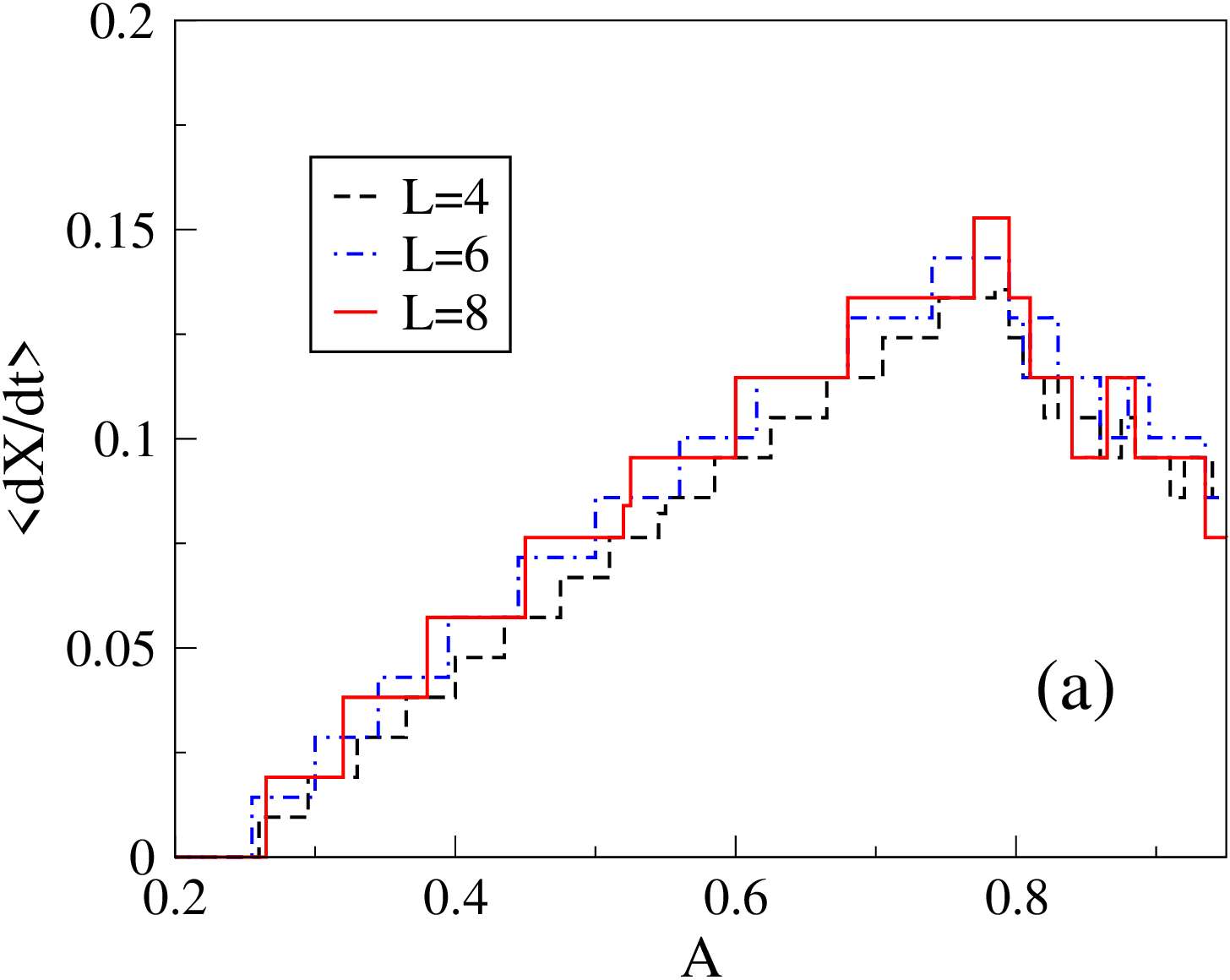}
\includegraphics[width=8.cm,height=8cm]
{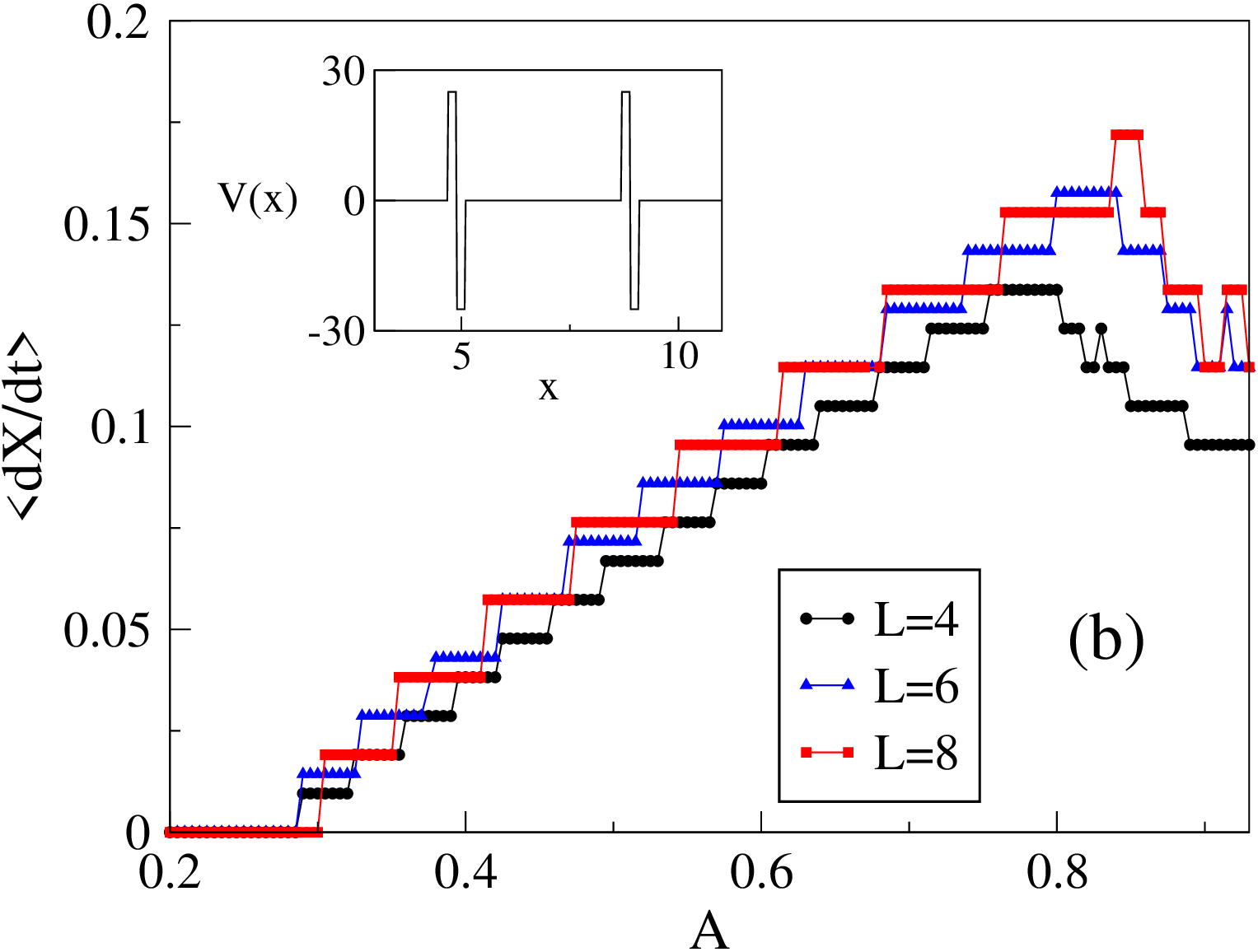}
\end{center}
\caption{Mean velocity vs. amplitude $A$ of the ac force. (a) Numerical solution
  of the CC-Eqs.
(\ref{ae}-\ref{ak}) with $U(X,l)$ given by Eq.(\ref{ar}). 
(b) Simulations of Eq.(\ref{aa})
  with the expression for $V(x)$ given by Eqs.(\ref{ab}) and (\ref{ap}). $V(x)$ is shown
  in the inset. 
The parameters used are
$\beta=1$, $\omega=0.015$, $h=25$, $b=0.1$, $x_{1}=0.8$, $x_2=1.0$ }
\label{Fig4}
\end{figure}

Our approach was to construct an array of inhomogeneities which produces an effective potential whose shape is approximately 
equivalent to the shape of $U_{opt}$, i.e. it must exhibit the two essential
features found in  $U_{opt}$ in the previous section. The required effective
potential can be obtained in four steps (here we skip working with $U(X)$ from
the 1-CC theory and proceed directly with $U(X,l)$ in Eq.(\ref{ar})):

1.) A single box inhomogeneity produces the single peak (\ref{aq}), two boxes with equal width $(b_1 = b_2 = b)$ and equal 
height, but {\em opposite} signs $(h_1 = h, h_2 = -h)$, produce the desired double peak structure, if they are placed at positions 
$x_1, x_2$ not too far separated.\\
2.) As to the second feature of $U_{opt}$, the long flat regions between the double peaks can easily be obtained by 
choosing a large period $L$, compared to the width of the double peaks. For the widths discussed below, a good range 
is $L = 4 - 8$; here $\langle v \rangle $ depends only weakly on $L$.\\
3.) We need to make the negative slope within a double peak very steep, and the positive slope between double peaks 
very weak, because these are features of $U_{opt}$. Both features can be achieved by making the two boxes narrower and
 higher (keeping the area fixed). Unfortunately the effect of this procedure is limited by the fact that the two peaks $U_1^{(0)}$
and $U_2^{(0)}$ which are produced by the two boxes, have a minimum width of about $1$ (for $l \simeq {\cal O} (1))$, see Eq. (\ref{aq}).
 This width is practically the same as that of the $sech^2$-peak produced by a $\delta $-function inhomogeneity, see 
below Eq. (\ref{aq}).\\
4.) Because of the limitations of step 3, we position the two boxes closer
together. This would make the negative slope steeper, if 
there were not a partial compensation of the positive and 
negative contributions $U_1^{(0)}$ and $U_2^{(0)}$ which reduces the heights of the two peaks. One can  
compensate this effect by further increasing the height $h$ of the boxes and
thus the peak heights. The optimal double-peak 
structure is produced by two very narrow boxes which are as close together as possible, namely touching each other 
without overlapping (e.g., $x_1 = 0.8, x_2 = 1.0, b = 0.1)$. Then the height can be increased as long as 
$v_{max}$ is reached for  $A \le 1$ (for the above example $h = 25$, see
Fig. \ref{Fig4}). There is no value in increasing the amplitude of the driving
force beyond $A \simeq 1$ because in the simulations for the original PDE 
(\ref{aa}) kink-antikink pairs may then appear spontaneously and degrade the ratchet
effect (see also end of section II).

The above four steps together yield an enormous gain for $v_{max}$ in the order of $300 \%$: In Fig. \ref{Fig4}a the maximal value 
of $\langle v \rangle$ is $0.1528$ for $\omega = 0.015$ , which has to be compared with the best value $0.0382$ for the case 
of three $\delta $-function inhomogeneities with the same $\omega $ \cite{luis04}. The value 
$0.1528$ for  $v_{max}$ is indeed quite high taking into account that this is
an average over positive and negative velocities, cf. Eq. (\ref{as}). Even
larger values of $v_{max}$ can be obtained by choosing a small damping
parameter, but we do not consider this regime for the reasons given at the end of
section II.

The above pair of very narrow and very high boxes that touch each other produces an effective potential which can be 
approximated very well by that resulting from a $\, \delta '$-function inhomogeneity. This can be shown by considering in Eq. (\ref{ar}) 
the limit $b \rightarrow 0, h \rightarrow \infty$ with finite $\epsilon:=(2b)^2 \cdot h$ for a box pair at $x_1$ and $x_2 = x_1 + 2 b$ in cell 
$n = 0$. The limit yields
\begin{equation}    
\label{ba}
U^{(0)} (X, l) = - \frac{4 \epsilon}{l} \,\tanh \frac{X - x_1}{l} \,\,\text{sech}^2 \frac{X - x_1}{l}
\end{equation}
which is identical with $U^{(0)}$ stemming from a $\delta '$-function inhomogeneity $g_1 = \epsilon \, \delta ' (x - x_1)$.

\section{Comparison with simulations}

In order to check the predictions of the above collective coordinate approach with
the optimized effective potential constructed in the previous section, we have
performed simulations for the full sine-Gordon system (\ref{aa}) with the box
inhomogeneities that produce the above effective potential: two narrow and high 
boxes per cell which have opposite signs and touch each other. We have
numerically solved Eq.(\ref{aa}) using the Heun scheme \cite{miguel} which
yields the same results as the Strauss-V\'azquez scheme \cite{vaz}, but considerably
faster; another advantage of the Heun scheme is the  possibility to include 
thermal noise \cite{luispre}. The spatial and temporal integration steps were 
$\Delta x=0.025$ and $\Delta t=0.005$.  The spatial interval for the simulations
had a length of 190 units. The simulations were performed taking as initial
condition a kink soliton Eq.(\ref{ad}) at rest, which we allow to evolve. Then, after some transient time, 
we compute the mean velocity integrating over a period of time. This
process is repeated again by varying the amplitude of the force, thus
sweeping over the whole interval for the force amplitude. We have checked that
the computation of the mean velocity does not
change over several periods.

 The results
in Fig.\ref{Fig4}b show a very good qualitative agreement with the CC approach
 results
in Fig.\ref{Fig4}a. Interestingly the simulation results are more sensitive to
the choice of the cell length than the CC approach results: both the maximum
$v_{max}$ of the
average velocity and its position change more when $L$ is increased
from $4$ to $8$. The best value is $v_{max}=0.172$, which is $350 \%$ higher 
than the best result $0.0382$ for the
$\delta$ -function inhomogeneities \cite{luis04}. 
In this case the highest
value of the soliton speed $|v(t)|$ is about $0.5$ which has to be compared with the critical velocity $c=1$.  

\begin{figure}
\begin{center}
\includegraphics[width=8.cm,height=8cm]
{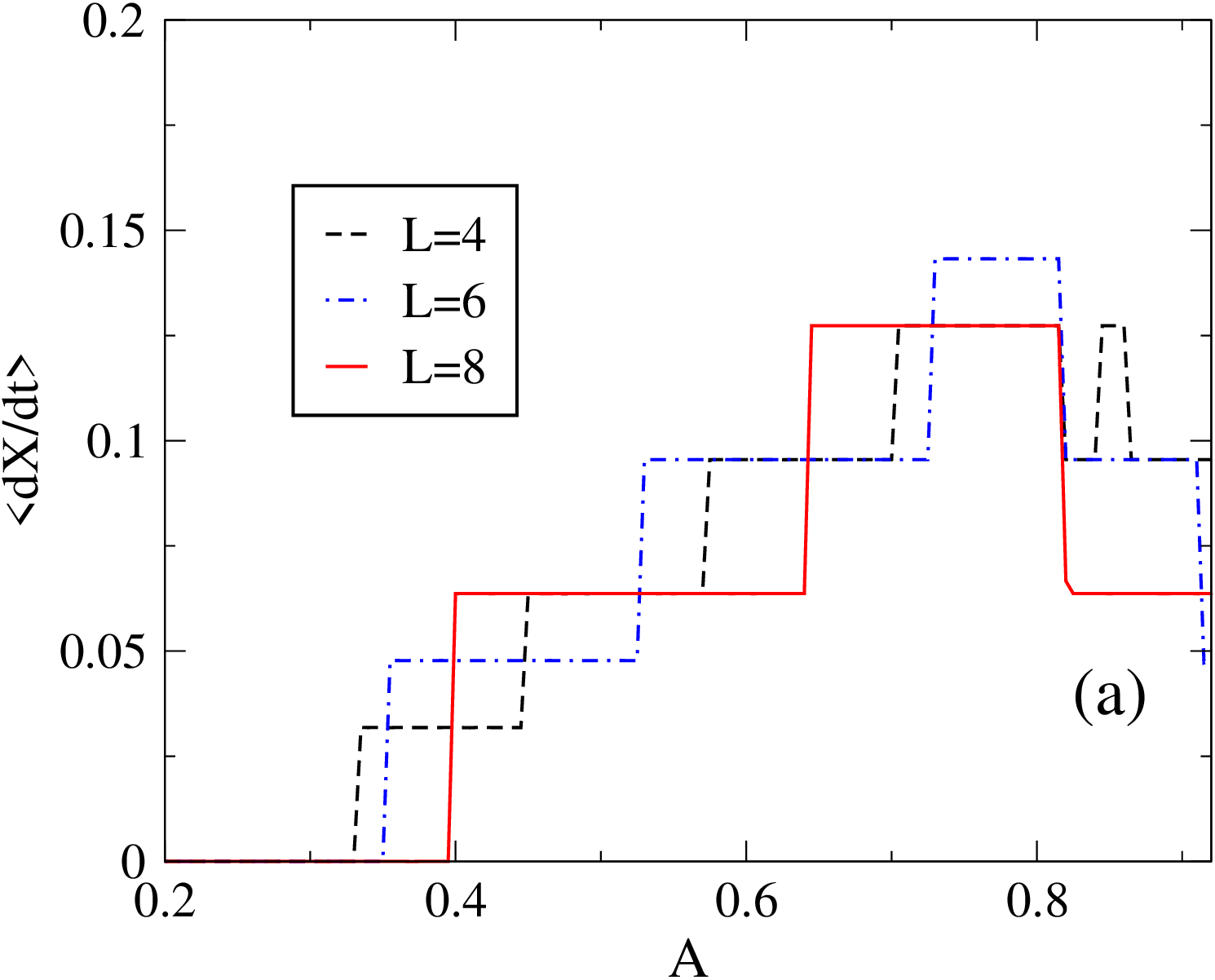}
\includegraphics[width=8.cm,height=8cm]
{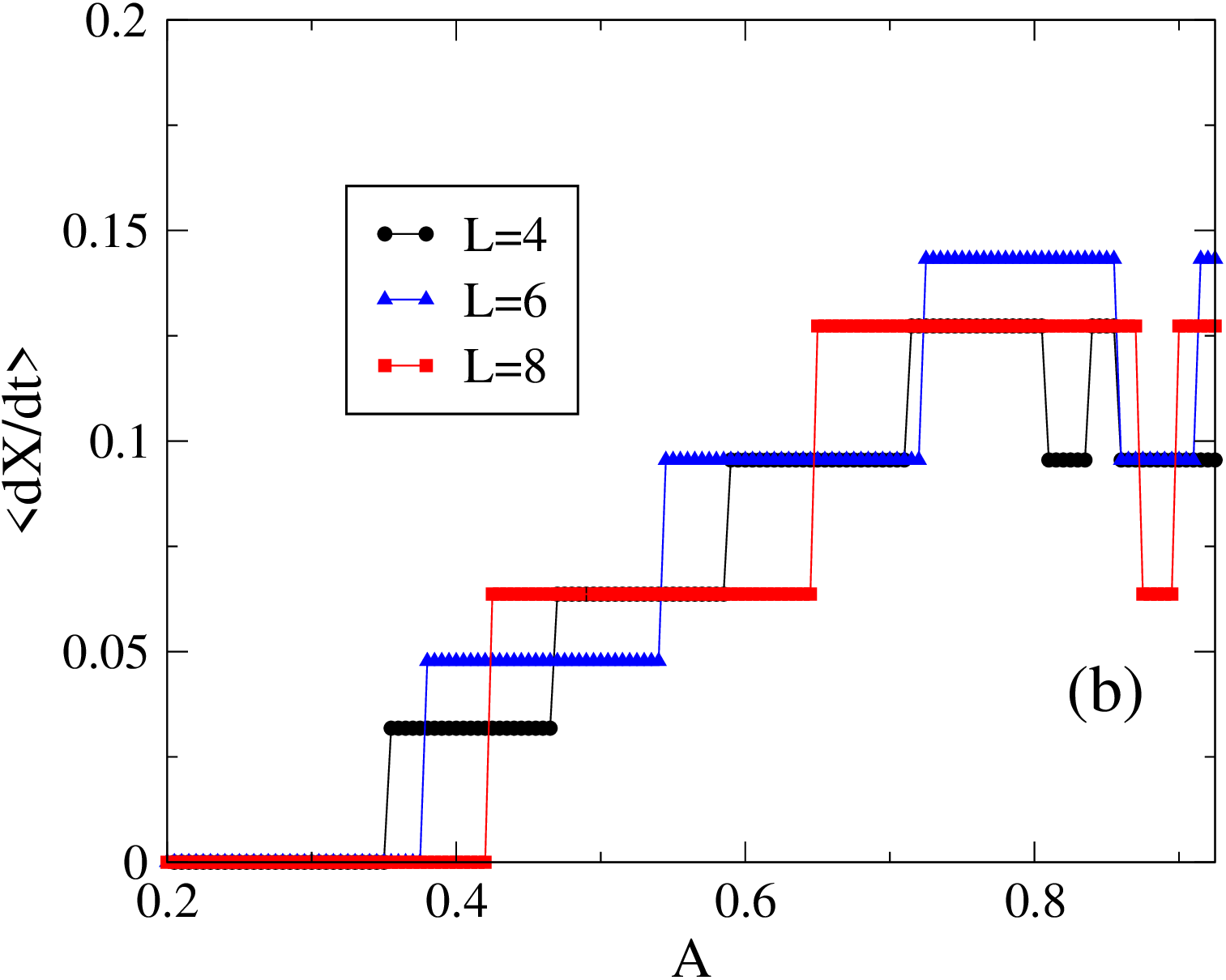}
\end{center}
\caption{Mean velocity vs. amplitude $A$ of the ac force. (a) Numerical
  solution of the CC-equations (\ref{ae})-(\ref{ak}). 
(b) Simulations of Eq.(\ref{aa})
  with the expression for $V(x)$ given by Eqs.(\ref{ab}) and (\ref{ap}).
The
  parameters used are $\beta=1$, $\omega=0.05$, $b=0.1$, $h=25$, $x_{1}=0.8$, $x_{2}=1.0$.
}
\label{Fig5}
\end{figure}

So far we have only considered the case of a low frequency ($\omega=0.015$) of
the driving force. Here both the $\delta$-function and the box inhomogeneities
produce a staircase in $\langle v \rangle$ vs. $A$. However, for higher frequencies the
situation changes: For $\omega=0.05$ the boxes still yield a staircase (Fig.\ref{Fig5})
while the $\delta$-functions yield several windows ( see case (b) in section
II). For $\omega=0.1$ (not shown) a 2-step staircase is seen for $L=4$, while two and one
windows are found for $L=6$ and $L=8$, respectively. 

\section{Summary and conclusions}

We have studied how ratchet transport in inhomogeneous sine-Gordon systems
can be strongly enhanced, proceeding in three steps:

1) In a collective coordinate approach the soliton dynamics was represented by
a relativistic particle ratchet with an effective potential $U$, which is
periodic and asymmetric. Using an ansatz in terms of Jacobi elliptic functions,
the optimal $U$
 for the particle ratchet was found. $U_{opt}$ exhibits essentially two
 features: a) double peaks consisting of a positive and a negative peak with a
 steep slope in between, b) long flat parts between the double peaks.

2) The inhomogeneities in the sine-Gordon system were chosen such that they
   produce an effective potential which exhibits the above features of
   $U_{opt}$. It turned out that a very good choice is a periodic array in
   which each cell contains two very narrow and high boxes with opposite
   signs that touch each other (this structure can be well approximated by a 
$\delta' $- function). The optimization of the parameters (width and height
   of the boxes and the cell length) yielded high values for $v_{max}$, the
maximum of the average soliton velocity $\langle v \rangle$ as a function
of the driving amplitude $A$. For the frequency range $[0.015,0.1]$
   $v_{max}$ is in the order of $0.15$, where the soliton speed $|v(t)|$
   reaches values which are not far away from the critical velocity $c=1$. For
   frequencies higher than $\omega=0.2$ the ratchet effect vanishes.

3) Simulations for the sine-Gordon system with the above inhomogeneities
   produce results for $\langle v \rangle$  vs. $A$ which show a very good
   qualitative agreement with the results of the collective coordinate
   approach.

From the results which we have obtained, we propose the following experiments on long Josephson junctions: Narrow 
inhomogeneities of opposite signs can be built by microresistors (critical
   Josephson current $J_{c}$ decreased) and microshorts ($J_{c}$ increased). So
   far only equidistant arrays of either type of inhomogeneities were used,
   but we do not see problems in producing arrays in which microresistors are placed
   close to microshorts.

A further enhancement of $v_{max}$ could be achieved by a replacement of the
sinusoidal driving force $f(t)$ by a more general periodic force (for
homogeneous sine-Gordon systems an ansatz in terms of Jacobi elliptic
functions, see below Eq. 25, yielded $v_{max}=0.16$ and $0.17$ in simulations
and a CC-approach, respectively \cite{proce}). Work on the joint optimization
of the inhomogeneities and the driving force is in progress.
   
\begin{acknowledgments}
F.G.M. thanks for the hospitality of the Theoretical Division and Center for
Nonlinear Studies at Los Alamos National Laboratory during his sabbatical
stay. Work at Los Alamos is supported by the USDoE. Travel between Spain and Germany was supported by the Ministerio de
Educaci\'on y Ciencia (Spain) and DAAD (Germany) through "Acciones Integradas
Hispano-Alemanas", HA2004-0034 and D/04/3997. A.S. is supported by the
Ministerio de Educaci\'on y Ciencia (Spain) under grants FIS2004-1001 and 
NAN2004-9087-C03-03, and by Comunidad de Madrid (Spain) under grants
UC3M-FI-05-007 and SIMUMAT-CM.

The authors also thank
N.R. Quintero (Sevilla) for the preprints of her work and fruitful discussions.
\end{acknowledgments}


\begin{thebibliography}{88}

\bibitem{Hanggi} P. H\"anggi and R. Bartussek, in {\it Nonlinear Physics of Complex Systems -- Current Status and Future Trends}, edited by J. Parisi, S. C. M\"uller, and W. Zimmermann (Lecture Notes in Physics No. 476, Springer, Berlin, 1996).

\bibitem{Astumian} R.D. Astumian and P.\ H\"anggi, Physics Today  {\bf 55}  (11), 22 (20002).

\bibitem{Ajdari} F.\ J\"ulicher, A.\ Ajdari, and J.\ Prost, Rev.\ Mod.\ Phys.\ {\bf 69}, 1269 (1997).

\bibitem{Reimann} P. Reimann, \ Phys.\ Rep. {\bf 361}, 57 (2002).

\bibitem{Linke} H.\ Linke, editor, {\em Ratchets and Brownian Motors:
  Basics,Experiments and Applications}, Appl.\ Phys.\ A {\bf 75} (special
  issue).

\bibitem{Zapata} I. Zapata, J. Luczka, F. Sols, and P. H\"anggi,\ Phys.\ Rev.\ Lett.\ {\bf 77}, 2292 (1996). 

\bibitem{Marchesoni} F. Marchesoni, \ Phys.\ Rev.\ Lett.\ {\bf 77}, 2364 (1996)

\bibitem{falo} F. Falo, P. J. Mart\'inez, J. J. Mazo, and S. Cilla, \ Europhys.\  Lett.\ {\bf 45}, 700 (1999).
\bibitem{trias} E. Tr\'ias, J. J. Mazo, F. Falo, and T. P. Orlando, Phys.\ Rev.\ E.\ {\bf 61},  2257 (2000).

\bibitem{salerno2} M. Salerno and N.R. Quintero,\ Phys.\ Rev.\ E.\ {\bf 65},  025602 (2002)

\bibitem{const} G. Constantini, F.\ Marchesoni, and M. Borromeo,  Phys.\ Rev.\ E.\ {\bf 65},  051103 (2002).

\bibitem{gorbach} A.V. Gorbach, S. Denisov, and S. Flach, Chaos {\bf 16}, 023126 (2006).

\bibitem{flach} S. Flach, Y. Zolotaryuk, A.E. Miroshnichenko, and  M.V. Fistul,  Phys.\ Rev.\ Lett.\ {\bf 88}, 184101 (2002).

\bibitem{Salerno-mixing} M. Salerno and Y. Zolotaryuk,\ Phys.\ Rev.\ E {\bf65}, 056603 (2002).

\bibitem{luisprl} L. Morales-Molina, N. R. Quintero, F. G. Mertens, and A. S\'anchez, Phys. Rev. Lett. {\bf 91}, 234102-1 (2003).

\bibitem{luis05} L. Morales-Molina, N. R. Quintero, A. S\'anchez and
  F. G. Mertens, Chaos {\bf 16}, 013117 (2006). 

\bibitem{JJ2}A.\ V.\ Ustinov, C. Coqui,A. Kemp, Y. Zolotaryuk, and M. Salerno, Phys.\ Rev.\ Lett.\ {\bf 93}, 087001 (2004).

\bibitem{Chacon} R. Chacon and N. R. Quintero, arXiv:physics/0503125 (2005).

\bibitem{luis04} L. Morales-Molina, F. G. Mertens, and A. S\'anchez,
  Eur. Phys. J. B {\bf 37}, 79 (2004).

\bibitem{luispre} L. Morales-Molina, F. G. Mertens and A. S\'anchez,
  Phys. Rev. E {\bf 72}, 016612  (2005).

\bibitem{usti} A. V. Ustinov, Phys. Lett. A {\bf 136}, 155 (1989).

\bibitem{waves} M. Reimoissenet, {\em Waves Called Solitons} (Springer, Berlin, 1999).
\bibitem{parrondo} J.M.R. Parrondo and B.J.Cisneros, Appl.Phys.A {\bf 75}, 179
  (2002).

\bibitem{goldobin} M. Beck, E. Goldobin, M. Neuhaus, M. Siegel, R. Kleiner,
  and D. Koelle, Phys. Rev. Lett. {\bf 95}, 090603 (2005).

\bibitem{rice}  M.J. Rice and E.J. Mele,\ Solid \ State \ Commun. {\bf 35}, 487 (1980); M.J. Rice, Phys.\ Rev.\ B {\bf 28}, 3587 (1983).

\bibitem{quintprl} N.R. Quintero, A. S\'anchez, and F.G. Mertens,Phys.\ Rev.\ Lett. \ {\bf 84}, 871 (2000).

\bibitem{quintpre} N.R. Quintero, A. S\'anchez, and F.G. Mertens, \ Phys.\
  Rev.\ E {\bf 62}, 5695 (2000).

\bibitem{wolfram} Mathematica, 5.0 (Wolfram Research Inc., Champaign, IL, 2003).   
\bibitem{lindner} B. Lindner and L. Shimansky-Geier, \ Phys. \ Rev. \ Lett. \
  {\bf 89} 230602 (2002). 

\bibitem{schreier} M. Schreier, P. Reimann, P. H\"anggi, and E. Pollak,\ Europhys.\ Rev.\ Lett.\ {\bf 44}, 416 (1998). 

\bibitem{miguel} M. San Miguel and R. Toral, Stochastic effects in physical systems
in {\em Instabitities and Nonequilibrium structures}, VI, E. Tirapegui, J. Mart\'inez
and R. Tienmann, eds Kluwer Academic Pub. 35-130 (2000).

\bibitem{vaz} W.\ A.\ Strauss and L.\ V\'azquez, J.\ Comput.\ Phys. {\bf 35},
61 (1990).

\bibitem{proce} R. Chacon and N.R. Quintero, proceedings of BIOCOMP 2005, to be published.

\end{thebibliography}
\end{document}